\documentstyle[preprint,aps]{revtex}

\newcommand{\Od}{{\cal O}}

\newcommand{\NP}[1]{ Nucl.\ Phys.\ {\bf #1}}
\newcommand{\ZP}[1]{ Z.\ Phys.\ {\bf #1}}

\newcommand{\PL}[1]{ Phys.\ Lett.\ {\bf #1}}

\newcommand{\PR}[1]{Phys.\ Rev.\ {\bf #1}}
\newcommand{\PRL}[1]{ Phys.\ Rev.\ Lett.\ {\bf #1}}

\begin{document}

\title{ Applicability constraints of the Equivalence Theorem}
\draft
\author{ A. Dobado$^{2}$ and J. R. Pel\'aez$^{3}$}
\address{Departamento de F\'{\i}sica Te\'orica.\\
Universidad Complutense.
28040 Madrid, Spain.}

\vskip .5 cm

\author{ M. T. Urdiales$^{2}$}
\address{
Departamento de F\'{\i}sica Te\'orica.\\
Universidad Aut\'onoma. 28049 Madrid. Spain.}
\date{February 1997}
\maketitle
\footnotetext[2]{electronic address:dobado@eucmax.sim.ucm.es}
\footnotetext[3]{electronic address:pelaez@eucmax.sim.ucm.es}
\footnotetext[4]{electronic address:mayte@delta.ft.uam.es}
\begin{abstract}
In this work we study the applicability of the
Equivalence Theorem, either for unitary models or
within an effective lagrangian approach.
There are two types of limitations:  the existence of a
validity energy window and the use of the lowest order
in the electroweak constants.
For the first kind, we consider some methods,
based on dispersion theory or the large $N$ limit, that allow
us to extend the applicability. For the second, we have obtained
numerical estimates of the effect of neglecting higher orders
in the perturbative expansion.
\end{abstract}

\section{Introduction}

Probably the main challenge for high energy physics  in the
next decades will be to establish the nature of the Symmetry Breaking Sector
(SBS) of the Standard Model (SM). The SBS
can either be weak or strongly interacting. In the first case, it is expected
that one or more Higgs bosons will be found at energies below 1 TeV.
In contrast, if the SBS is strongly interacting, new states are expected
above that energy, mostly in form of resonances.

In this work we will only deal with an strongly interacting SBS,
whose experimental signature will be a production enhancement
of longitudinally polarized gauge bosons
($V_L$), at energies higher than
1 TeV \cite{Ve77,LeQu77}.
 The link between $V_L$'s and the SBS is
the so called Equivalence Theorem (ET)
\cite{LeQu77}-\cite{HeKu92}, which 
relates amplitudes containing $V_L$'s with
those amplitudes where every $V_L$ is 
replaced by its associated Goldstone Boson
(GB). Indeed, the theorem is used in two ways:
to connect the hidden SBS with physical observables but also to
simplify calculations, since
GB are scalars, and hence much easier to handle. It is therefore quite
usual to evaluate the GB amplitudes and then translate them 
using the ET.

Based only on the SBS group structure,
it is possible to derive some identities known
as Low Energy Theorems (LET) \cite{ChGo87},
 which predict the behavior of GB at very low energies,
irrespective of the precise mechanism that yields the symmetry breaking.
During the last few years, a very powerful formalism
\cite{DoHe89} has been developed
in order to deal with the strongly interacting SBS in a model independent way.
Such a description is inspired in Chiral Perturbation Theory (ChPT)
\cite{We67}, which
has been remarkably succesful in describing low energy pion physics.
The idea is to build
an effective lagrangian as a derivative expansion, whose
first order term (with two derivatives) reproduces the LET.
The next order includes operators up to dimension four \cite{ApBe80}
and encodes the details of the undelying SBS in a set of parameters.
Just by changing the values of these parameters, we can
differentiate alternative
SBS models \cite{DoHe89,RES} (like a heavy-Higgs SM \cite{HeRu94},
or a QCD-like model \cite{EsRa90}). It is expected that
some of them will be measured at future
accelerators like LHC or NLC \cite{DoHe95}.

However, the fact that  chiral amplitudes are obtained
as an expansion in powers of $p/(4\pi v)$, where 
$v\simeq250\mbox{GeV}$ and $p$ is the external momenta, spoils the
usual proof of the ET \cite{Ve90}. Only very recently it has
been derived a new version of the ET for 
effective lagrangians \cite{HeKu94}-\cite{DoPe94}.
Nevertheless, the lack of unitarity of chiral amplitudes apparently
imposes severe constraints on the validity of the theorem.
The purpose of this work is to study in detail
such constraints, how they affect the usefulness
 of the ET approximation and how
they could be avoided.

The  plan of the paper is as follows. In section 2 we
state the ET with a brief sketch of its derivation.
Then, in section 3, we identify the approximations
performed to obtain the theorem. In section 4 we study 
the case of $V_L$ scattering and,
in section 5, we analyze some unitarization procedures that
extend the ET applicability range. In section 6 we present
some brief conclussions.

\section{The Equivalence Theorem}

As it is well known, 
the GB and the $V_L$ are
related through a gauge transformation. Indeed, the GB, 
denoted as $\omega^a$, disappear from
the lagrangian
if we use a unitary gauge. However, most of the calculations are performed in
the so called $R_\xi$ or renormalizable gauges, where the gauge fixing term
has the following generic form:
\begin{equation}
{\cal L}_{GF}=-\frac{1}{2}\left( \partial^\mu V_\mu^a + \xi M \omega^a
\right)^2
\end{equation}
where M is the mass of $V^a$. The gauge fixing term is nothing
but a delta function in functional space and thus, intuitively
it is telling us, in momentum space, that
$V_\mu^a p^\mu/M \sim \omega^a$. But, as far as,
\begin{equation}
\epsilon_L^\mu = \frac{p^\mu}{M} + v_\mu \quad \mbox{with} \quad
v_\mu = \Od\left( \frac{M}{E}\right)
\label{v}
\end{equation}
we can replace longitudinal gauge bosons by
their associated GB at energies $E\gg M$. 
Naively, that is the Equivalence Theorem
\cite{LeQu77,CoLe74}.

Rigorously,
we have to reformulate the ET within Quantum Field Theory in terms
of amplitudes or S matrix elements. The first formal
proof of this kind was given by Chanowitz and Gaillard in \cite{ChGa85}
in the context of the Minimal Standard Model (MSM).
The complete derivation of the ET is rather complicated and
we will not go through it in here, although in Fig.1 we have
included a simple
sketch of the most important steps.
 Indeed, they used
the Slavnov-Taylor identities which are obtained from the BRS invariance
of the quantum lagrangian, which now has the role that gauge invariance played
in the classical reasoning. That proof was later simplified in
\cite{GoKo86}.
We say it is formal because one also has to take into account
renormalization  and some correction factors $K$ 
have to be included in the  final ET statement \cite{YaYu88}.
These factors have been calculated in the most usual renormalization
schemes as well as in those where their value is just one \cite{HeKu92}.
The Slavnov Taylor identities yield relations among
renormalized Green functions which, using the Reduction Formula,
can be translated to amplitudes.
Once all these is done, the momentum factors can be replaced  using
Eq.\ref{v}. After a little bit of algebra
\cite{ChGa85,DoPe94} it is possible to write separately
all the powers of $v_\mu$. That expression is nowdays known as the Generalized
Equivalence Theorem (GET) and it reads
 \begin{eqnarray}
T(V^{a_1}_L,...,V^{a_n}_L;A) = \sum_{l=0}^{n} (-i)^{l}
\left( \prod_{j=1}^{l}K^{a_j}_{\alpha_j} \right)
\left( \prod_{i=l+1}^{n}v_{\mu_i} \right)
\bar T(\omega_{\alpha_{1}} ...\omega_{\alpha_l},
V^{\mu_{l+1}}_{a_{l+1}}... V^{\mu_n}_{a_n};A)
\label{GET}
\end{eqnarray}

This formula deserves some comments (see \cite{DoPe94} for details):

\begin{itemize}

\item $A$ stands for any set of other
{\em physical} fields i.e., not $V_L$'s.

\item On the
 left hand side we have the amplitude we were looking
for, containing an arbitrary number $n$ of $V_L$'s.
On the right hand side we have a sum over all amplitudes $T$ where we have
replace one, two, ... up to $n$
$V_L$'s
by their associated GB, 
$\omega^\alpha$. The bar over the amplitudes  means that
we are considering all 
permutations of indices. Observe that
the only energy dependent factors that multiply the
amplitudes are the $v_\mu$,
one per each $V_L$  which has not been substituted by its
GB.

\item The GB  can be written in any parametrization. That is why
the $K$ correction factors have two indices, 
since the $V^a_L$ carry indices of
$SU(2)_L\times U(1)_Y$, whereas the $\omega^\alpha$ belong to the
broken quotient space. Thus, the
$K$ factors also choose the right GB combination that is "eaten"
by each $V^a_L$.

\end{itemize}

From this expression it is fairly simple to obtain the ET
using Eq.\ref{v} and the fact that unitarity constrains 
how fast the amplitudes can
grow with the energy. In the limit $E\gg M$,:
\begin{equation}
T(V^{a_1}_L,...,V^{a_n}_L;A)\simeq
\left(\prod_{j=1}^{n}K^{a_j}_{\alpha_j} \right)
T(\omega_{\alpha_1} ...\omega_{\alpha_n};A) + \Od\left(\frac{M}{E}\right)
\label{ET}
\end{equation}
That is the Equivalence Theorem in its most common version, but
unfortunately it is not valid when dealing with chiral
Lagrangians. It is easy to understand why:
on the one hand, the
ET is a high energy limit, since we need $E\gg M\simeq gv$
in order to neglect the
$v_\mu$ terms in Eq.\ref{v}. On the other hand, the chiral approach is a
low-energy limit, and we are expanding the amplitudes as follows:
\begin{equation}
T(V_{\mu_1}^{a_1}
... V_{\mu_l}^{a_l},\omega_{\alpha_{l+1}} ...\omega_{\alpha_n};A)
\simeq
\sum_{k=0}^{N} a_{l}^k \left(\frac{E}{4\pi v} \right) ^k + \sum_{k=1}^{\infty}
a_{l}^{-k} \left( \frac{M} {E}\right)^{k}
\label{Champ}
\end{equation}
(we do not display the $\mu_i$ and $a_i$ indices in the
$a^k_l$ factors). Thus, we are taking a high and low energy
limit at the
same time, which may be possible or not.

Nevertheless the proof in Fig.1, can be modified
for the effective formalism: First, considering the nonlinear character
of the symmetry, which only affects the $K$ factors
 \cite{HeKu94}. Second, taking into account that chiral amplitudes
grow with the energy, as in the above equation.
As a consequence, 
the $\Od(M/E)$ terms contained in the $v_\mu$
factors mess up with the chiral $E/4\pi v$ powers.
Fortunately, between the scale $4\pi v$ and 
$M\sim g v$ there is a $g/4\pi$ factor,
which finally allows us to extract the leading order 
in $g$ of every coefficient of the chiral expansion.
Thus, the ET reads
\begin{eqnarray}
T(V^{a_1}_L,..., V^{a_n}_L;A)  &\simeq&
 \left(\prod_{j=1}^{n}K^{a_j (0)}_{\alpha_j} \right)
T_L^N(\omega_{\alpha_1}...\omega_{\alpha_n};A)  \nonumber\\
\label{ETefflag}
&&+\Od\left( g \mbox{ or } g' - \mbox{supressed} \right)
+\Od\left(\frac{M}{E}-\mbox{supressed} \right)
+\Od\left( \frac{E}{4\pi v} \right)^{N+1}
\end{eqnarray}
where $T_L^N$ is the $\Od(p^N)$ chiral amplitude at lowest
order in the electroweak couplings $g$ or $g'$. Note that
the $K$ factors  have also been expanded and their
only contribution comes from the lowest order in $g$ or $g'$,
denoted by $K^{(0)}$.

The ET may have not changed very much in its form, but
we have to neglect more terms, which will limit the validity of the
whole approach. To which extent these approximations are consistent 
and useful
will be the topic of the next sections.

As a last remark, we also want to point out that,
very recently, it has appeared a new derivation of the ET, based
on Lorentz invariance \cite{HeKu95}, which reaches similar results
for both cases, although
it clarifies the interpretation of the ET in different Lorentz frames. 
Here it has been assumed that we are at the CM in a
collision where all initial or final $V_L$'s have comparable momenta,
generically $\Od(p_i)=\Od(E)=\Od(\sqrt{s})$.
Notice that we have also been considering $\Od(M_W)=\Od(M_Z)$.

\section{Applicability constraints}

Up to the moment, the ET has been considered as a formal
mathematical statement. We now want to know when we are
allowed to neglect all the higher order terms that appear in its expression.
From the previous section,
there are two kind of restrictions: those related with
the energy regime, and those connected with neglecting higher orders in the
electroweak expansion. We will address them separatedly.

\subsection{Conditions due to the energy expansion}

As we have already seen, there are two formulations of the ET and each one
is used depending on the unitarity properties of the amplitudes.
As a consequence we have two sets of applicability conditions:

\begin{itemize}

\item For {\bf models respecting unitarity.-} That is,
 for instance, the case of the MSM.
As far as the amplitudes cannot grow as a power of the
energy, we can extract the ET leading contribution by simply neglecting
the $v_\mu$ terms in Eq.\ref{v}. Thus we can use the first formulation of
the ET in Eq.\ref{ET} and we only have to ask for $M\ll E$.

\item When dealing with {\bf effective lagrangians.-} The
amplitudes are obtained as truncated series up to a
given power $(E/4\pi v)^N$.
Hence, in order to neglect higher $E/4\pi v$ terms we have to
demand that $E\ll 4\pi v$. But we still want to approximate $p/M\sim\epsilon_L$
to obtain the ET, so that we need
\begin{equation}
M\ll E\ll 4\pi v=\frac{8\pi}{g} M
\label{AC1naive}
\end{equation}
But that is not all in
Eq.\ref{ETefflag}, we are neglecting the
$\Od(M/E-\mbox{supressed})$ terms while keeping those of $\Od(E/4\pi
v)^N$. Thus, we also need
\begin{equation}
\Od\left(\frac{M}{E}-\mbox{suppressed}\right)\ll
\Od\left( \frac{E}{4\pi v} \right) ^N
\label{AC2naive}
\end{equation}
Notice that the last condition is more strict
for higher $N$. 
\end{itemize}
 Nevertheless, even in unitary models like the MSM, amplitudes
are obtained perturbatively.  Then it could be possible 
that some high order terms could also be hidden by the $\Od(M^2/E^2)$
neglected contributions \cite{HeKu95}. 
In that case we would have similar conditions
to those in \ref{AC2naive}. In this work we will only study such conditions
within the chiral formalism.

\subsection{Conditions due to the expansion in $g$ and $g'$.}

It has been shown that the ET only yields the lowest 
order in the $g$ expansion within the effective formalism. 
But even in the unitary cases it is still very
frequent to work at lowest order in the electroweak couplings $g$
and $g'$, since then there are no internal lines of gauge bosons
and one does not have to deal with complicated propagators inside
loops. Once more we have to demand that the terms we drop should be
smaller that those we keep, that is
\begin{equation}
\Od\left(g \mbox{ or } g' -\mbox{supressed}\right)\ll
\Od\left( \frac{E}{4\pi v} \right) ^N
\label{AC3naive}
\end{equation}

In practice, if one wants to work at a higher
$g$ or $g'$ order, one has to calculate:

\begin{itemize}

\item [a)] The amplitude $T(\omega,...\omega)$ to next order.

\item [b)] The $K$ factors, whose expressions are rather complicated but
that have been obtained to one loop in several renormalization  schemes
\cite{HeKu92}.

\end{itemize}

Within the chiral formalism, we are using Eq.\ref{ETefflag} 
and thus, in order to work at
higher orders of $g$ or $g'$,
it is not enough with steps a) and b) but we also have to
obtain:

\begin{itemize}
\item [c)] The lowest order contribution in the electroweak couplings
of $T(V_L,\omega,...\omega)$. In other words, we need the amplitudes
where one of the longitudinal gauge bosons has not been replaced by its
corresponding GB (Indeed, we have to calculate the amplitudes with the
$V_L$ in all the different positions).

\end{itemize}

This complication shows up in the last step of Fig.1
because the above amplitudes appear in the Generalized Equivalence
Theorem multiplied by just one $v_\mu$ factor (for details see
\cite{DoPe94}). Notice that for our purposes,
$v_\mu$ counts as $\sim g\Od(v/E)$. Indeed, 
we have explicitly checked at tree level that without this
last contributions one does not get the same results on both sides
of the ET. Very recently it has been published a work \cite{EsMa95}
where these corrections have been included, again at tree level,
for the $W^+W^+\rightarrow W^+W^+$ process and they have found a
perfect agreement. Besides, they have obtained the $K$
renormalization factors in the on-shell scheme.

\subsection{Special Kinematic Regions}

The constraints in the previous sections are obtained 
on very general grounds. However, the
numerical values of the coefficients of the chiral expansion
also determine the size of the different contributions.
If they become to big or too small they can spoil the previous 
naive counting. Indeed that is the 
case in some special kinematic regions.

\begin{itemize}

\item It is well known
that the exchange of massless particles, like photons or GB,
in  $t$ or $u$ channels
could lead to infrared divergencies at low scattering angle $\theta$.
If those singularities appear in the terms we are neglecting,
then the ET would be a very bad approximation, although the
formal mathematical statement will still hold.
When dealing with $V_LV_L\rightarrow V_LV_L$, 
those divergencies coming from photon
exchange, are at least $\Od(e^2)\sim\Od(g^2)$. As a consequence,
if we just keep the
lowest order in $g$, we would be droping the divergent (dominant)
term at low scattering angle, and in that region the
approximation will be completely inadequate.
It is always possible however, to keep precisely that part
of the $\Od(g^2)$ contributions which is $\Od(e^2)$. 
Mathematically, it may seem at first
not very consistent to keep just part of the $\Od(g^2)$ contributions,
but physically it can be accepted, as far as it is the dominant part
at low $\theta$ \cite{HeKu96}.

\item It could also happen that in a given channel the term obtained
from the ET may be proportional to, let us say, $u$ \cite{HeKu96}.
It is then possible to make that term as small as we want by approaching
$\theta\rightarrow\pi$. Therefore, even though the mathematical
expression of the ET is still correct, we are not allowed to neglect
the $\Od(M/E)$ term. 

\end{itemize}

\section{Physical processes}

The above constraints on the
applicability of the ET can be relaxed for most
of the processes that have been 
proposed to probe the SBS in future colliders:
\begin{eqnarray*}
V_LV_L\rightarrow V_LV_L  \quad \stackrel{ET}{\longleftrightarrow} \quad \omega
\omega
\rightarrow \omega \omega
\\ \nonumber
q q'\rightarrow V_LV_L  \quad \stackrel{ET}{\longleftrightarrow} \quad q q'
\rightarrow \omega \omega
\\ \nonumber
\gamma \gamma \rightarrow V_LV_L  \quad \stackrel{ET}{\longleftrightarrow}
\quad \gamma  \gamma
\rightarrow \omega\omega 
\end{eqnarray*}
As far as only an even number of $V_L$ appear,
the expressions of the amplitudes will only  contain
{\it even powers} of momenta or the energy.
Consequently: 

\begin{itemize}

\item We just need $M^2\ll E^2$, instead of $M\ll E$.

\item When dealing with the effective lagrangian formalism we need ($N=2n$)
\begin{eqnarray}
&&M^2 \ll E^2 \ll (4 \pi v)^2  \label{AC1} \\
&&\Od\left(\frac{M^2}{E^2} -\mbox{supressed}\right)\ll
\Od\left( \frac{E}{4\pi v} \right) ^N \label{AC2}\\
&&\Od\left(g \mbox{ or } g' -\mbox{supressed}\right)\ll
\Od\left( \frac{E}{4\pi v} \right) ^N \label{AC3}
\end{eqnarray}

\end{itemize}

In order to obtain numerical values for the above applicability limits,
we have to keep in mind
that the numbers that we will give are not strict mathematical bounds, but
rather some rough estimates. 

We will then start with the lowest energy bound, which is common to both
formulations of the ET, namely, $M^2\ll E^2$.
If we demand, for example, $E^2$ to be one order of magnitude bigger than 
$M^2$,
that means that $E>300$ GeV. 

Within the effective lagrangian formalism we also have the $4\pi v$ 
upper bound,
 but in practice that is too optimistic. One possible  way to estimate the
upper validity range, is to compare it with that of ChPT
  and pion physics. As a
matter of fact, that is an analogous 
formalism to the one we are using in this
work, although it is scaled down to energies of the order of 1GeV. More
precisely, in ChPT, the scale $4\pi f_\pi\simeq 1 \mbox{GeV}$ plays the very
same
role as $4\pi v\sim 3\mbox{TeV}$ here.
The $\Od(p^4)$ ChPT amplitudes are known to work
reasonably well only up to $500 \mbox{MeV}\simeq 2\pi f_\pi$
 in the best cases \cite{We67}. Therefore, we should not trust the
chiral formalism beyond energies of the order of $2\pi v\simeq 1.
5\mbox{TeV}$. In brief, in a first approximation we find that
the applicability of the ET within the $\Od(p^4)$
chiral lagrangian formalism is
restricted to the following energy range: 
\begin{equation}
0.3 \mbox{TeV} < E <  1.5 \mbox{TeV},
\end{equation}
maybe less. Indeed, in some models, we expect resonances to appear 
below $1.5$ TeV, and fo that the chiral approach is not valid. 
Notice that the high energy constraint {\em is inherent to the 
effective lagrangian approach}, it is not really due to the usage of
the ET.

Concerning Eqs.\ref{AC2} and \ref{AC3}, they have been 
obtained by demanding that the
neglected terms should be smaller than the 
effects we want to observe.
Therefore, Eqs.\ref{AC2} and \ref{AC3} 
are telling us which energies we have to reach
in order to see the desired $\Od(p^N)$ effects, but below that energy
the rest of the amplitude will be enough to describe the relevant physics.
These constraints are much more process dependent than the others above, 
and we will study them for $V_LV_L\rightarrow V_LV_L$.

\subsection{Scattering of longitudinal gauge bosons}

This process is one of the most interesting at LHC
since one of the signatures of
an stronly  interacting SBS is the enhancement of $V_L$ interaction. 
Within the chiral lagrangian formalism,
the first term, which is $\Od(E^2/v^2)$, is universal 
and it just reproduces the LET.
Therefore, if we want to differentiate between alternative 
SBS, we have to look at the next order, which, in general is
$\Od(E^2/v^2\cdot E^2/(4\pi v)^2)$. Hence Eq.\ref{AC2}
can we written as:
\begin{equation}
\frac{M^2}{E^2} \ll \frac{E^4}{v^2(4\pi v)^2}  \Rightarrow E > 4\pi v
\sqrt[3]{\frac{g}{32\pi^2}} \simeq 0.4 \mbox{TeV}
\end{equation}
But we are also neglecting a g-suppressed term in the tree amplitude,
which is $\Od(E^2/v^2\cdot M^2/E^2)$, and thus  Eq.\ref{AC3} yields:
\begin{equation}
\frac{M^2}{v^2} \ll \frac{E^4}{v^2(4\pi v)^2}  \Rightarrow E > 4\pi v
\sqrt[2]{\frac{g}{8\pi}} \simeq 0.5 \mbox{TeV}
\end{equation}
(Note that if we do not keep track of the $2$ in $M=gv/2$, the lower 
bound is $0.7$ TeV). 
This lower bound does {\it not} mean that the whole
$\Od(p^4)$ amplitude cannot be
used with the ET below 0.5 TeV. It is just telling us that the $\Od(p^4)$
terms will only be seen above that energy.
Notice however, that the $\Od(p^4)$ amplitude also contains $\Od(p^2)$ terms,
whose consistency condition \ref{AC3}  reads:
\begin{equation}
\frac{M^2}{v^2} \ll \frac{E^2}{v^2}
 \Rightarrow M^2\ll E^2 
\end{equation}
which is again the usual constraint of the ET. 
These $\Od(p^2)$ terms dominate the chiral expansion at low
energies, so that they are enough to obtain the correct result between 0.3 and
0.5 MeV, but they do not differentiate alternative SBS.

To summarize, within the effective lagrangian formalism and for
$V_L$ elastic scattering,  we can expect the ET to be
a reasonable approximation when 0.3 TeV $< E <$ 1.5 TeV, although the
truly $\Od(p^4)$ operators can only be tested
 whenever:
\begin{equation}
 0.5 \mbox{TeV} \le E \le 1.5 \mbox{TeV}
\end{equation}

Notice that the high energy bound is only due to the
use of the effective formalism. 
This restriction is very dissappointing since the enhancement 
of the $V_L$ interaction is
expected to be bigger at higher energies.
Moreover, one of the main features of strongly interacting
systems are resonances, which are expected to appear around 1 TeV.
Indeed, the existence of such resonant states implies the
saturation of unitarity bounds and a breakdown of the chiral expansion.
Thus, in some cases, the 1.5 TeV bound could be
even very optimistic
(as for instance for the MSM with a heavy Higgs, that could appear as a very
broad scalar resonance at energies near 1 TeV).
For all these reasons, it would be very
interesting to extend the applicability of the ET to much higher energies.
We will refer to this possibility in section 5.

Let us remark once again that the numbers in the above equations should be understood as orders of magnitude, since we do not know what are the actual values of the chiral parameters and the coefficients in the chiral expansion.

Nevertheless, there are some analysis of the LHC capabilities to measure the
$\Od(p^4)$ parameters, although they have been performed just at tree
level with the $\Od(p^2)$ and the $\Od(p^4)$ lagrangian. It seems likely
to determine their values within the order of magnitude
expected for a generic strong  SBS \cite{DoHe95}. These analysis do not
make use of the ET, but it could be very helpful in case the whole one-loop
calculation was performed.

\subsubsection{Numerical estimates.}

We have seen that for longitudinal gauge boson scattering,
the biggest terms that we are neglecting seem to be those of $\Od(g^2)$.
However, their size has been estimated comparing energy powers, although it
could well happen that the actual values of the coefficients
in the expansion may spoil the power counting arguments.
With that purpose, we have calculated the amplitudes of several
processes with and without the ET, by considering
all the graphs obtained, at tree level,
from both the $\Od(p^2)$ and $\Od(p^4)$ terms in the
Electroweak Chiral Lagrangian. On the one hand,
the amplitude containig external
GB has only been calculated at lowest order in $g$ and $g'$, since
that is how the ET should be applied. On the other hand,
the amplitude obtained without
the ET contains the higher order $g$ and $g'$ contributions.
In such way, by comparing both amplitudes, we get an estimate of the error
caused when using the ET.

The results can be found in Fig.2, where we have plotted the cross sections
for $ZZ\rightarrow ZZ, W^+W^-\rightarrow ZZ$ and
$W^\pm Z\rightarrow W^\pm Z$. The solid lines represent the calculation
with the ET and the dashed lines without it. Their difference is only due to
higher order $g$ and $g'$ effects, and that we have checked by letting
$g,g'\rightarrow 0$. In that limit, the two lines superimpose.

Let us now remember that the point in using 
chiral lagrangians
is to reproduce, within the same formalism, any symmetry breaking
mechanism that follows the SM symmetry breaking pattern. That is enough
to fix the $\Od(p^2)$, but the operators at  $\Od(p^4)$, whose form
is also dictated by symmetry requirements, are all affected by
some multiplying parameters, usually denoted as $\alpha_i$ \cite{ApBe80}. 
For our purposes we are only interested in
$\alpha_5$ and $\alpha_4$, since they are the only ones appearing in the
above amplitudes.
Their values
depend on the underliying symmetry breaking mechanism and, by varying
them, we are able to mimic any strong SBS. We
have indeed performed
our calculations for many different models, 
but in Fig.2 we have only displayed
the two which are usually present in the literature:
\begin{itemize}
\item The SM with a heavy Higgs (on the left column),
 whose values for the $\alpha_i$
can be obtained by matching the Green functions of the Higgs model
with those coming from the effective lagrangian approach \cite{HeRu94}.
At tree level, the matching yields: $\alpha_5=v^2/8M_H$ and
$\alpha_i=0$ if $i\alpha_i\neq 5$.
Therefore, by choosing $M_H\simeq 1 \mbox{TeV}$, we have
$\alpha_5\simeq0.008$

\item A QCD-like model (on the right column), whose estimated values for the
chiral parameters are: $\alpha_5=-0.0016$ and $\alpha_4=0.0008$.
They have been obtained from a large-$N_c$ expansion of the QCD
 effective action \cite{EsRa90}.

\end{itemize}

Observe that, in Fig.2, we have represented the SM-like model only up
to 1 TeV, since that is the mass we have chosen for the Higgs and therefore
we expect it to show up as a resonance at that energies. Such resonant
states cannot be reproduced within the chiral approach,
although it can be noticed that there is a huge increase in
the cross section. The QCD-like model cross sections have been plotted
up to 1.5 Tev; that is the energy up to where
the chiral approach works for QCD, but rescaled to the electroweak SBS.

In Fig.2 it can be noticed that the typical cross sections for
longitudinal gauge boson scattering is of the order of 1 nb. In contrast
the contributions from higher orders in $g$ or $g'$ are typically of the
order of 0.1 nb. Only in those channels which almost vanish,
these corrections may be important. In those channels with a significant
signal, the effects of neglecting them are just corrections
of the order of $\simeq 10 \%$.

\section{Dispersion relations}

Up to this moment we have seen that the most severe constraint to
the applicability of the ET appears when it is applied within
the effective lagrangian approach. Even though the $\Od(g,g')$ effects
may be negligible, the lack of unitarity strictly limits how high in energies
we can trust the whole approach. Indeed, the most typical feature
of strong interactions is the appearence of resonances, which are
deeply related to the saturation of unitarity, where we do not expect
the effective approach to work properly.

Our purpose in this section is not to review all the unitarization methods
proposed in the literature, but only those
that have been succesfully tested with pion data in the framework of ChPT.
Notice that QCD has the
very same symmetry breaking pattern,
$SU(2)_L\times SU(2)_R\rightarrow SU(2)_{L+R}$,
that we have in the electroweak SBS. Indeed, there is a similar
chiral lagrangian formalism for pions, but rescaled from
$v\simeq250 \mbox{GeV}$ down to $f_\pi\simeq 92\mbox{MeV}$.
The only differences are the actual values of the $\alpha_i$ parameters,
the small masses of the pions (which are pseudo-GB) and the fact that
we do not couple electroweak gauge bosons.
However, we can use the ET to translate the rescaled pion amplitudes into
longitudinal gauge boson amplitudes.
That is why we use pion physics as a reference.
Apart from that, we also look for methods that do not require the explicit
introduction of  resonances as additional fields. That may be useful
 in pion physics since the resonances are already known, but that is not the
case of the Electroweak SBS. Thus we have restricted ourselves to
illustrate the dispersive approach, since it is able to unitarize the
chiral amplitudes and reproduce resonances just from the knowledge of
their low-energy behavior.

The general idea is to reproduce the unitarity cut which is present when we
extend the partial waves of a given process to the complex $s$ plane.
This cut is due to the existence of a threshold. In addition,
the partial waves also have
a left cut due to crossing symmetry and possibly some poles in the
second Riemann sheet, which are closely related to the resonances.
If we apply Cauchy's Theorem to this analytic structure, we can obtain the
value of the amplitude at any $s$ in terms of integrals of the
imaginary part of the amplitude over the cuts,
plus maybe some polynomials of the energy.
The low-energy behavior is given by the polynomials,
which can be approximated
using the chiral approach. The integrals carry the high energy and resonant
behavior although they also correct
the bad high energy behavior of the polynomials.
The interesting point is that sometimes some of these integrals can
 be calculated exactly.

\subsection{The Inverse Amplitude Method}

In the elastic scattering of GB we have partial waves
of definite (weak)-isospin I and angular momentum J. In the chiral approach
they are obtained as $t_{IJ}=t_{IJ}^{(0)}+t_{IJ}^{(1)}+ \Od(s^3)$.
The LET are reproduced by $t_{IJ}^{(0)}$,
which is an $\Od(s)$ polynomial. Formally, $t_{IJ}^{(1)}$
can be seen as an $\Od(s^2)$ polynomial whose coefficients
can contain logarithmic terms which provide
a first approximation of the cuts.
The unitarity constraint,
 for real values of $s$ above threshold (the elastic cut),
is nothing but:
\begin{equation}
\mbox{Im} t_{IJ}=\sigma \vert t_{IJ} \vert^2
\label{uni}
\end{equation}
 The
$\sigma$ factor is the GB two body phase space and for
pure GB is 1 (for pions, which are massive, it is
$\sigma(s)=\sqrt{1-4m_\pi^2/s}$). Obviously the polynomials
cannot satisfy this quadratic constraint and
they violate unitarity.
Nevertheless, the chiral amplitudes satisfy unitarity
 {\em perturbatively}:
$\mbox{Im} t_{IJ}^{(1)}=\sigma \vert t_{IJ}^{(0)} \vert^2$.

The important remark is that we can obtain
 the imaginary part of the inverse
amplitude exactly: $\mbox{Im} 1/t_{IJ}=\sigma$. Thus, it
is possible to calculate the dispersive integral
of $1/t$ over the elastic cut and then solve for $t$ \cite{DoPe92}.
The result is \cite{DoHe90}:
\begin{equation}
t_{IJ}=\frac{t_{IJ}^{(0)}}{1-t_{IJ}^{(1)}/t_{IJ}^{(0)}}
\label{IAM}
\end{equation}
This formula presents several interesting properties:
\begin{itemize}
\item It can be easily checked that the above formula satisfies the
unitarity constraint in eq.\ref{uni}.
\item If we expand it again at small $s$,
we recover the original chiral amplitude. Thus, at low energies
we have not spoiled the good behavior given by the effective lagrangian.
\item It is able to accomodate poles in the second Riemann sheet
and reproduce resonances.
\end{itemize}

Indeed, it has been shown that the Inverse Amplitude Method (IAM) is
able to reproduce the $\rho(770)$ resonance in $\pi\pi$ scattering
\cite{DoHe90},
and the $K^*(892)$ in $\pi K$ scattering \cite{DoPe92}.
Their correct masses and witdths
are obtained with chiral parameters which are compatible
with those coming from a fit of pure ChPT to the low energy data.
As it should be, the resonances always appear related to poles in
the second Riemann sheet, by the formula
$\sqrt{s_{pole}}\simeq M+\Gamma/2$, where $M$ and $\Gamma$ are,
respectively, the mass and width of the resonance \cite{DoPe92}.

The IAM was first applied to longitudinal gauge boson scattering
 in \cite{DoHeTru90}. In Figure 3, we are showing
the results as continuous lines. We represent the phase shifts $\delta_{IJ}$
for the three lowest angular momentum channels:
$(I,J)=(0,0),(1,1),(2,0)$.
The dashed lines correspond to the pure chiral lagrangian approach
with the ET, and, as we have already seen, should not be trusted above
1.5 TeV or even before if there is a resonance.
 For illustrative purposes,
we have chosen two typical scenarios for the electroweak SBS:

\begin{itemize}
\item Those graphs on the left have been obtained
with the chiral parameters that mimic the SM. As we commented before,
they are estimated from a matching of the SM Green functions
 and those coming from the chiral lagrangian \cite{HeRu94}.
 This time,
however, we are calculating at one loop and thus,
for $M_H=1.2\mbox{TeV}$,
we have to use $\alpha_5=0.0045$ and $\alpha_4=-0.0020$.
Notice that the expected scalar resonance in
a heavy-Higgs scenario does indeed appear, in contrast with
the non-unitarized approach. As a matter of fact,
we have chosen $M_H$ so that the mass of the
scalar resonance is $1\mbox{TeV}$.

\item Those graphs on the right column are just a rescaled version
of pion-pion scattering in the limit $m_\pi=0$. Therefore
they represent a QCD-like SBS. The parameters we have used
are obtained directly from the IAM applied to elastic pion
scattering data: $\alpha_5=-0.00103$ and $\alpha_4=0.00105$ \cite{DoPe92}.
Once again, and in contrast to
the pure effective approach, there is
 a $\rho$-like resonance in the $(1,1)$ channel.

\end{itemize}

\subsection{Inelastic Case}

Dispersive techniques can also be applied to the inelastic case
\cite{DoPe93}
or to calculate form factors of inelastic processes
\cite{Tru88}.
As far as the reaction contains two GB, we can expect
strong rescattering effects to become relevant, and then
unitarization would come into play. We will illustrate such effects
using the $\gamma\gamma\rightarrow Z_LZ_L$ process, since the
dispersive unitarization techniques have also been succesfully applied
to $\gamma\gamma\rightarrow \pi^0\pi^0$. We refer to \cite{DoPe93}
for further details. We can therefore aaply the very same techniques
of that work since both reactions are related
through the ET in the $m_\pi=0$ once we rescale $f_\pi\simeq 92\mbox{MeV}$
up to $v\simeq 250\mbox{GeV}$.

The dashed curves in Figure 4 correspond to the
cross section of the reaction calculated within the
pure effective formalism \cite{HeRu92}. This process is forbiden
at tree level,
and the one-loop contributions become dominant.

As far as the effective
amplitude does not depend on $\alpha_i$, the predictions
of the effective approach are the same for every underlying SBS.
However, by looking
at the continuous lines, which correspond to the unitarized
calculation, we see how the cross section differ from a
SM-like to a QCD-like model. The unitarization has been performed
by imposing the elastic phase shifts of Figure 3 as
the phases of the $\gamma\gamma\rightarrow V_L V_L$ amplitudes.
For the absolute value of the amplitude we keep the pure result
from the effective lagrangian. In the SM case, it can be
noticed that the corrections become important as soon as the
resonance appears, and they modify the result at the qualitative level.

\subsection{The large-$N$ limit}

Finally, we will also like to comment that there is another approach
that yields unitary amplitudes,
which has also been tested with pion data and ChPT.
The idea is to take the large-$N$ limit \cite{Cas84},
 $N$ being the number of GB,
and keep the dominant terms in a $1/N$ expansion. The resulting
amplitudes do not behave as polynomials in the
energy and they satisfy unitarity up to $\Od(1/N)$.
  Therefore it is possible to use them together with the
simplest version of the ET, eq.\ref{ET}, which has less severe
applicability bounds and allows us to trust the calculations at
higher energies. This approximation is specially well suited for the scalar
channel (but not so well for
the others since they are subdominant and would require higher orders
in $1/N$). Indeed, it has been shown that the large-$N$ approximation
is able to reproduce an scalar resonance in a heavy-Higgs model, together
with its associated pole in the second Riemann sheet \cite{DoPe96}.

\section{Conclussions}

Usually, the ET has been applied to the SM or other unitary models.
Except in special kinematic regions, once the 
renormalization effects are taken
into account, it is enough to demand that $E\gg M_W$.

In contrast, when dealing with non-unitary models like in the
electroweak chiral approach, further limitations appear. First, there is
the typical constraint $E\ll 4\pi v$, of the effective lagrangian formalism.
But apart from that, the version of the ET valid in this case
does not include higher order electroweak
corrections. Even more, if one is interested in probing the $\Od(p^N)$
terms of the amplitudes, there is an additional condition:
\begin{equation}
\Od\left( \frac{M}{E}, g \mbox{ or } g' - \mbox{supressed}\right)
\ll \Od\left( \frac{E}{4\pi v}\right)^N
\end{equation}

In this work we have obtained, on the one hand, some estimates
of the above constraints for the processes
of interest in future accelerators. It seems that the ET
applicability window to test the truly $\Od(p^4)$
(model dependent) contributions is roughly: 0.5TeV $\leq$ E $\leq$
1.5TeV, {\em if there are no resonances in that range}, otherwise
it will be much smaller.
On the other hand, we have obtained, from detailed numerical
calculations, that the effect of neglecting higher order
electroweak corrections is typically $\simeq 10\%$, in the cross sections
of the relevant channels.

Therefore, it seems that neglecting the electroweak
effects will not be very relevant when applying the ET.
In contrast, the constraints on the energy range do
severely limit the simultaneous application of
the ET and the chiral approach. Such applicability range
can be considerably enlarged using dispersive techniques, or non-perturbative
chiral expansions (like the large-$N$ limit), which improve
the unitarity behavior of the amplitudes, without changing
their low-energy properties.

In conclusion, and in view of our results, it
seems that the simultaneous application of the ET
and the effective
chiral approach to the physically relevant cases is severely
limited unless it is complemented with some non-perturbative techniques.

\section*{Acknowledgments}

J.R.P would like to thank the Theory Group at Berkeley
for their kind hospitality, the Jaime del Amo fundation for financial
support as well as M.J.Herrero and H.J.He
for helpful discussions on the subject.
This work has been partially supported by the Ministerio de
Educaci\'on y Ciencia (Spain) (CICYT  AEN93-0776).

\newpage

\centerline{\LARGE FIGURE CAPTIONS}

\paragraph{Figure 1.-} The proof of the Equivalence Theorem.

\paragraph{Figure 2.-}  The cross-sections
in this figure have been
obtained at tree level from ${\cal L}_2$
and ${\cal L}_4$. With different
values of the $\alpha_i$ parameters in ${\cal L}_4$
we mimic either a Heavy Higgs model (left column) or a
QCD-model (right column). The dashed lines are the
calculation without the ET. The solid lines are obtained
from the ET. Hence they are calculated at lowest order in $g$.
The differences between both lines are therefore
numerical estimates
of the effect of neglecting higher order electroweak
corrections.

\paragraph{Figure 3.-} Phase shifts $\delta_{IJ}$
for $V_LV_L\rightarrow V_LV_L$ scattering. The pictures on
the left correspond to the heavy Higgs SM-like case and those on the right
to the QCD-like model. The dashed lines are the predictions of the effective
lagrangian and the solid lines are the unitarized results. The black dots
stand at $\delta_{IJ}=90^o$ pointing the existence of a resonant state.

\paragraph{Figure 4.-} Cross sections for
$\gamma\gamma\rightarrow Z_LZ_L$ both for a heavy Higgs SM, or a QCD-like
model. The dashed lines come from the effective lagrangian approach
and the solid lines are the unitarized results.


\begin{thebibliography}{ }

\footnotesize

\bibitem{Ve77} M. Veltman,  Acta Phys. Pol. {\bf B8}, 475 (1977).
\bibitem{LeQu77} B. W. Lee, C. Quigg and H. Thacker, \PR{D16}, 1519 (1977).
\bibitem{CoLe74} J. M. Cornwall, D. N. Levin and G. Tiktopoulus,
\PR{D10},   1145 (1974).\\
C. E. Vayonakis,  Lett. Nuovo Cim. {\bf 17},  383 (1976).

\bibitem{ChGa85} M. S. Chanowitz  and M. K. Gaillard, \NP{B261},
   379 (1985).

\bibitem{GoKo86} G. K. Gounaris, R. Kogerler and H. Neufeld,
\PR{D34},   3257 (1986).

\bibitem{YaYu88} Y. P. Yao and C. P. Yuan, \PR{D38},   2237 (1988).\\
J. Bagger and C.Schmidt, \PR {D41},  264 (1990).

\bibitem{HeKu92} H. J. He, Y. P. Kuang and X. Li, \PRL {69},
  2619 (1992); \PR{D49}4842 (1994).\\
H.J.He and W.Kilgore, \PR{D55} 1515 (1997).

\bibitem{ChGo87} M. S. Chanowitz, M.  Golden and H. Georgi,
\PR{D36}, 1490 (1987).

\bibitem{DoHe89} A. Dobado and H. Herrero, \PL{B228}, 495 (1989).\\
J. Donoghue and C. Ramirez, \PL{B234}, 361 (1990).

\bibitem{We67} S. Weinberg, \PRL{19},   1264 (1967).\\
 J. Gasser and H. Leutwyler,   Ann. of Phys. {\bf 158},
 142 (1984).

\bibitem{ApBe80} T. Appelquist and C. Bernard \PR{D22} 200  (1980).\\
A. Longhitano, \PR{D22},  1166 (1980),  \NP{B188}, 118 (1981).

\bibitem{RES} J. R. Pel\'aez, \PR{D55} 4193(1997);  Proceedings of the 
DPF/DPB Summer Study on New Directions on High Energy Physics.
Snowmass 96.

\bibitem{HeRu94}
D. Espriu and M. J. Herrero \NP{B373}, 117 (1992).\\
M.J.Herrero and E. Ruiz Morales, \NP{B418}, 431 (1994).

\bibitem{EsRa90}D. Espriu, E. de Rafael and J. Taron,
 \NP{B345}, 22 (1990), Erratum, ibid. {\bf B355}, 278 (1991).

\bibitem{DoHe95} A. Dobado et al. \PL{B352}, 400 (1995).\\
A. Dobado and M. T. Urdiales, \ZP{C71}, 659 (1996).

\bibitem{Ve90} H. Veltman, \PR{D41}, 264 (1990).

\bibitem{HeKu94} H. J. He, Y. P. Kuang and X. Li,
\PL {B329}, 278 (1994). \\
C. Grose-Knetter and I. Kuss, \ZP {C66}, 105 (1995).

\bibitem{DoPe94} A. Dobado and J. R. Pel\'aez, \PL{B329}
(1994) 469. Addendum, ibid. {\bf B335}, (1994) 554;
 \NP{B425},  110 (1994). Errata, ibid. {\bf B434}, (1995) 475.

\bibitem{HeKu95} H. J. He, Y. P. Kuang and C.-P.Yuan,
\PR{D51} 6463 (1995).

\bibitem{HeKu96} H. J. He, Y. P. Kuang and C.-P.Yuan,
\PR{D51} 6463 (1995).

\bibitem{EsMa95} D. Espriu and J. Mat\'{\i}as, \PR{D52}, 6530 (1995).

\bibitem{DoPe92} A. Dobado and J. R. Pel\'aez, Phys. Rev.
 {\bf D47}, 4883 (1992); LBL-38645, hep-ph/9604416 to appear in \PR{D}.

\bibitem{DoHe90}  A. Dobado, M. J. Herrero and
T. N. Truong,  Phys. Lett. {\bf B235}, 134 (1990).

\bibitem{DoHeTru90} A. Dobado, M. J. Herrero and
T. N. Truong,  Phys. Lett. {\bf B235},  129 (1990).\\
A. Dobado, M. J. Herrero and J. Terron, \ZP{C50},  205 (1995); ibid 465.


\bibitem{DoPe93} A. Dobado and J. R. Pel\'aez,
 Z. Phys. {\bf C57}, 501 (1993).

\bibitem{Tru88} T. N. Truong,
\PRL{61} (1988) 2526, ibid {\bf 67},  2260 (1991).\\
 T. Hannah, \PR{D51} (1995) 103, \PR{D52},  4971 (1995).

\bibitem{HeRu92}  M. J.Herrero and E. Ruiz Morales, \PL{B296}
 397 (1992).

\bibitem{Cas84} R. Casalbuoni, D. Dominici
 and R. Gatto, \PL{B147}  419 (1984).  \\
 M. B. Einhorn, \NP{B246}  75 (1984). \\
A. Dobado and J. R. Pel\'aez,  Phys. Lett. {\bf B286}
 136 (1992).

\bibitem{DoPe96} A. Dobado, J. Morales, J. R. Pel\'aez and M. T. Urdiales,
\PL{B387}  563 (1996).

\end{thebibliography}
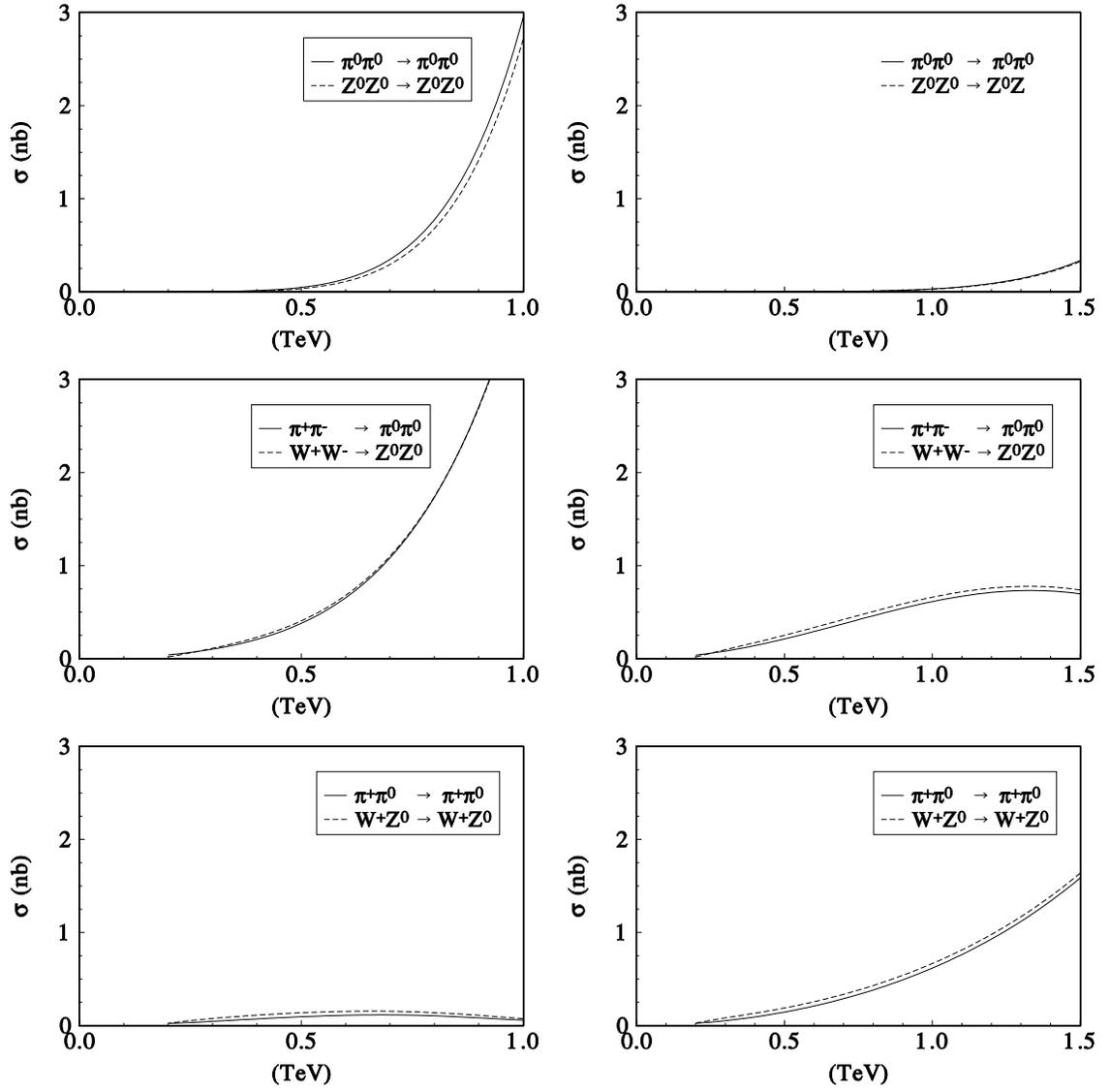
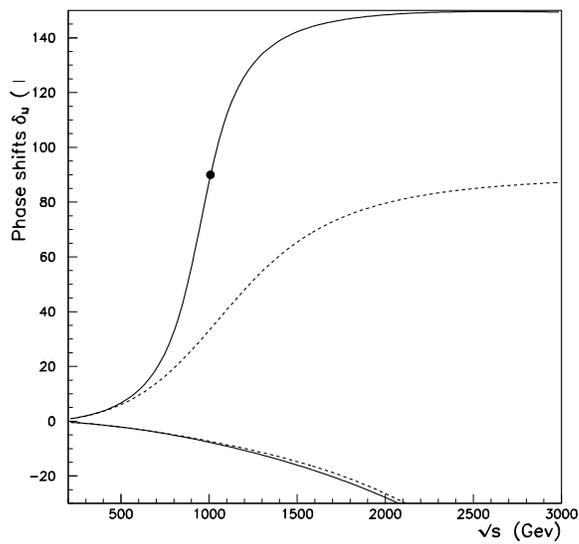
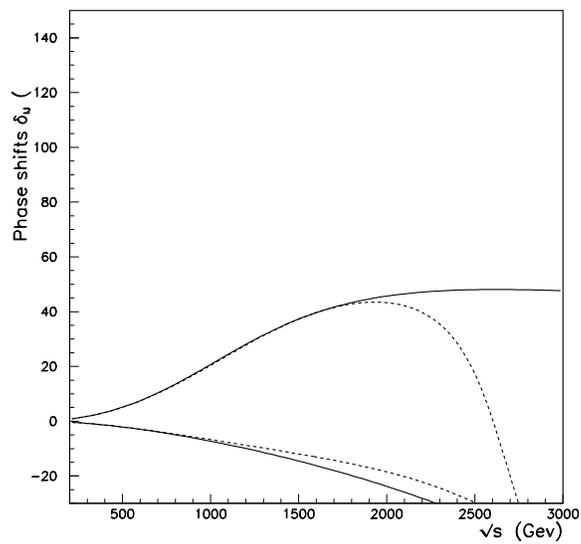
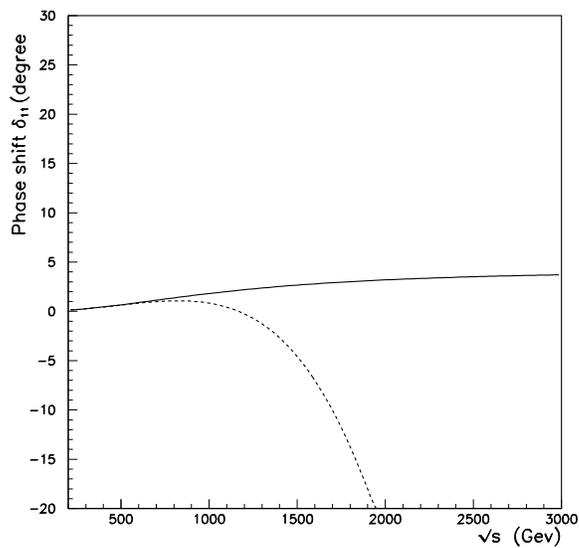
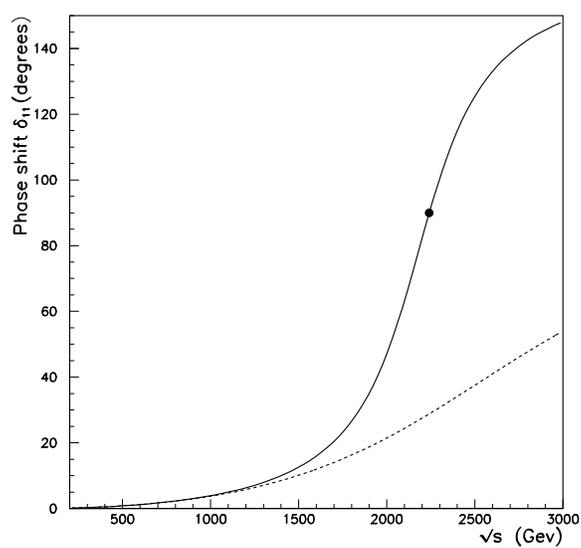
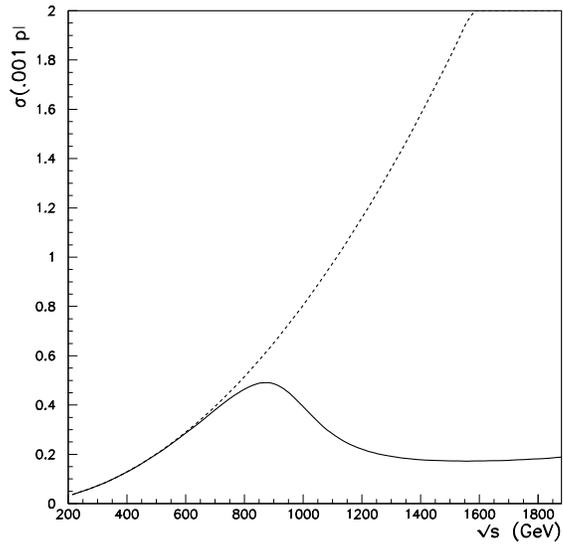
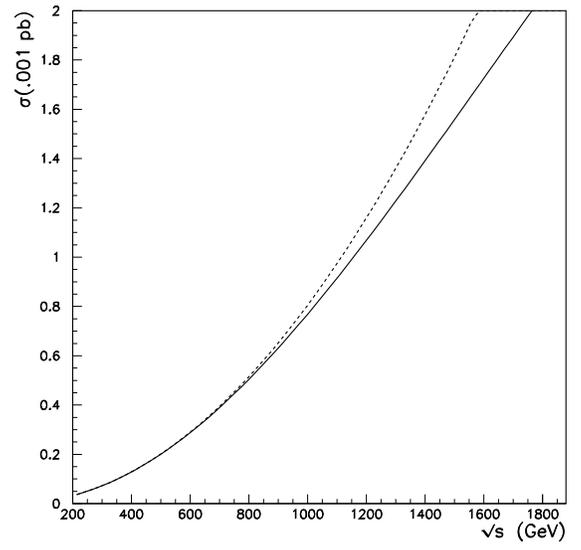
\end{document}